\documentclass[twocolumn,showpacs,preprintnumbers,amsmath,amssymb]{revtex4}
\usepackage{amsfonts}

\usepackage{graphicx}
\usepackage{dcolumn}
\usepackage{bm}
\usepackage{mathrsfs}
\usepackage{graphicx}
\usepackage{float}
\usepackage{amsmath}
\usepackage{subfigure}

\begin{document}

\title{  Quantum Key Distribution by Utilizing Four-Level Particles  }

\author{Tao Yan, Fengli Yan}

\email{flyan@mail.hebtu.edu.cn}

\affiliation {~\\
    College of Physics Science and Information Engineering, Hebei Normal University, Shijiazhuang 050016, China\\
   ~
Hebei Advanced Thin Films Laboratory,  Shijiazhuang 050016, China}

\date{January 4, 2010}

\begin{abstract}
{We present a  quantum key distribution protocol based on four-level
particles entanglement. Furthermore, a controlled quantum key
distribution protocol is proposed by utilizing three four-level
particles. We show that  the two protocols are  secure.
 }
\end{abstract}

\maketitle

Quantum key distribution (QKD) is one of the most important branches
of quantum cryptography, and plays an important role in perfectly
secure communication between two parties. In classical cryptography,
there is nothing to prevent an eavesdropper from monitoring the key
distribution channel without being detected by legitimate users. In
quantum cryptography, the principle of quantum mechanics was
introduced  to ensure the security of  the key distribution channel.
Since the seminal work of Bennett and Brassard \cite {BB84}, quantum
cryptography has been developed quickly \cite {Ekert91, BBM92,B92,
GinsinRevModPhys2002,LongLiu2002,DengLong2003,DengLong2004,Hwang2003,
Lo2005,DengLiZhouZhang2005,LiDengZhou2007,X.B.WangPRL2004,X.B.WangPR2004,X.B.WangPR2005}.
By using  Einstein-Podolsky-Rosen \cite {EPR} (EPR) correlations,
Ekert \cite {Ekert91} suggested a QKD protocol, in which one can
certify that the particles of EPR pairs are safely transmitted in
the quantum channel by using Bell's theorem \cite {Bell}. In 1992,
Bennent et al. \cite {BBM92} proposed a simpler  EPR protocol
without invoking the Bell's theorem.

In present paper we  put forward a KQD protocol based on the
four-level particle entanglement first. Then a controlled quantum
key distribution protocol is proposed by using the entangled three
four-level particles as the quantum key distribution channel.

Now, let us give the  quantum key distribution protocol by utilizing
two entangled four-level particles as the quantum channel.

Suppose that two legitimate correspondents, Alice and Bob, share  a
number of the following entangled quantum state
\begin{equation}\label{1}
\begin{array}{ll}
~&|\chi\rangle_{AB}\\
=&\frac {1}{2}(|01\rangle+|10\rangle-|23\rangle-|32\rangle)_{AB}\\
=&\frac
{1}{2}(|\psi^-\rangle|\phi^+\rangle+|\phi^-\rangle|\psi^+\rangle+|\psi^+\rangle|\phi^-\rangle+|\phi^+\rangle|\psi^-\rangle),
\end{array}\end{equation}
which is also called  the quantum channel. Here the four-level
particles $A$ and $B$ belong to Alice and Bob, respectively; the
states $|i\rangle$ $(i=0,1,2,3)$ stand for the four eigenstates of
four-level particles $A$ and $B$; the states $|\phi^\pm\rangle$,
$|\psi^\pm\rangle$  are well defined as follows
 \begin{equation}\label{2}
\begin{array}{ll}
|\phi^+\rangle=\frac {1}{\sqrt 2}(|0\rangle+|3\rangle), & |\phi^-\rangle=\frac {1}{\sqrt 2}(|0\rangle-|3\rangle),\\
|\psi^+\rangle=\frac {1}{\sqrt 2}(|1\rangle+|2\rangle), &
|\psi^-\rangle=\frac {1}{\sqrt 2}(|1\rangle-|2\rangle).
\end{array}
 \end{equation}
Obviously, the states in Eq.(2) are orthogonal to each other and
 constitute a basis of four-dimensional Hilbert space.

We introduce the following  operators
\begin{equation}\label{3}
\begin{array}{lll}
\sigma_x & = & |3\rangle\langle 0|+|0\rangle\langle 3|+|1\rangle\langle 2|+|2\rangle\langle 1|,\\
\upsilon_x & = & |2\rangle\langle 0|+|0\rangle\langle 2|+|3\rangle\langle 1|+|1\rangle\langle 3|,\\
\sigma_z & = & |3\rangle\langle 3|+|1\rangle\langle 1|-|0\rangle\langle 0|-|2\rangle\langle 2|,\\
\upsilon_z & = & |2\rangle\langle
2|+|3\rangle\langle3|-|0\rangle\langle 0|-|1\rangle\langle
1|.\\\end{array}
\end{equation}

Clearly,  $\sigma_x, \upsilon_x,
 \sigma_z, \upsilon_z$  are Hermitian operators and
\begin{equation}\label{4}
\begin{array}{l}
\sigma_x^2=\upsilon_x^2= \sigma_z^2=\upsilon_z^2= I.\end{array}
\end{equation}
Here $I$ is an identity operator on the four-dimensional Hilbert
space. So the eigenvalues of the operators $\sigma_x, \upsilon_x,
 \sigma_z, \upsilon_z$ can only be 1 or -1.

It is easy to prove that \addtocounter{equation}{1}
\begin{align}
\sigma_x^A\otimes\sigma_x^B|\chi\rangle_{AB}&=-|\chi\rangle_{AB}, \tag{\theequation .1}\\
\upsilon_x^A\otimes\upsilon_x^B|\chi\rangle_{AB}&=-|\chi\rangle_{AB}, \tag{\theequation .2}\\
\sigma_z^A\otimes\sigma_z^B|\chi\rangle_{AB}&=-|\chi\rangle_{AB}, \tag{\theequation .3}\\
\upsilon_z^A\otimes\upsilon_z^B|\chi\rangle_{AB}&=|\chi\rangle_{AB}.
\tag{\theequation .4}
\end{align}

Next we show that if a quantum state $|\Psi\rangle_{ABE}$ satisfies
\addtocounter{equation}{1}
\begin{align}
\sigma_x^A\otimes\sigma_x^B|\Psi\rangle_{ABE}&=-|\Psi\rangle_{ABE},\tag{\theequation .1}\\
\upsilon_x^A\otimes\upsilon_x^B|\Psi\rangle_{ABE}&=-|\Psi\rangle_{ABE},\tag{\theequation .2}\\
\sigma_z^A\otimes\sigma_z^B|\Psi\rangle_{ABE}&=-|\Psi\rangle_{ABE},\tag{\theequation .3}\\
\upsilon_z^A\otimes\upsilon_z^B|\Psi\rangle_{ABE}&=|\Psi\rangle_{ABE},\tag{\theequation
.4}
\end{align}
then the quantum state
$|\Psi\rangle_{ABE}=|\chi\rangle_{AB}|\alpha\rangle_E$. Here $E$
denotes the environment of the system consisted of particles $A$ and
$B$; $|\alpha\rangle_E$ is the quantum state of environment. Of
course, the eavesdropper is included in the environment.

Apparently, the most general state $|\Psi\rangle_{ABE}$ is of the
form
\begin{equation}
\begin{array}{lll}
&&|\Psi\rangle_{ABE}\\
&=&|00\rangle_{AB}|\alpha_1\rangle_E+|01\rangle_{AB}|\alpha_2\rangle_E+|02\rangle_{AB}|\alpha_3\rangle_E\\
&&+|03\rangle_{AB}|\alpha_4\rangle_E+|10\rangle_{AB}|\alpha_5\rangle_E+|11\rangle_{AB}|\alpha_6\rangle_E\\
&&+|12\rangle_{AB}|\alpha_7\rangle_E+|13\rangle_{AB}|\alpha_8\rangle_E+|20\rangle_{AB}|\alpha_9\rangle_E\\
&&+|21\rangle_{AB}|\alpha_{10}\rangle_E+|22\rangle_{AB}|\alpha_{11}\rangle_E+|23\rangle_{AB}|\alpha_{12}\rangle_E\\
&&+|30\rangle_{AB}|\alpha_{13}\rangle_E+|31\rangle_{AB}|\alpha_{14}\rangle_E+|32\rangle_{AB}|\alpha_{15}\rangle_E\\
&&+|33\rangle_{AB}|\alpha_{16}\rangle_E,
\end{array}
\end{equation}
where  $|\alpha_i\rangle_E$, $i=1,2,\cdots, 16$ stand for  the
unnormalized quantum states of the environment.

By considering  Eq.(6.1) we obtain
\begin{equation}
\begin{array}{ll}
|\alpha_1\rangle_E=-|\alpha_{16}\rangle_E,&|\alpha_2\rangle_E=-|\alpha_{15}\rangle_E,\\
|\alpha_3\rangle_E=-|\alpha_{14}\rangle_E,& |\alpha_4\rangle_E=-|\alpha_{13}\rangle_E,\\
|\alpha_5\rangle_E=-|\alpha_{12}\rangle_E,&|\alpha_6\rangle_E=-|\alpha_{11}\rangle_E,\\
|\alpha_7\rangle_E=-|\alpha_{10}\rangle_E,&|\alpha_8\rangle_E=-|\alpha_9\rangle_E.\\\end{array}
\end{equation}
So Eq.(7) becomes
\begin{equation}
\begin{array}{ll}
&|\Psi\rangle_{ABE}\\
=&(|00\rangle-|33\rangle)_{AB}|\alpha_1\rangle_E+(|01\rangle-|32\rangle)_{AB}|\alpha_2\rangle_E\\
&+(|02\rangle-|31\rangle)_{AB}|\alpha_3\rangle_E+(|03\rangle-|30\rangle)_{AB}|\alpha_4\rangle_E\\
&+(|10\rangle-|23\rangle)_{AB}|\alpha_5\rangle_E+(|11\rangle-|22\rangle)_{AB}|\alpha_6\rangle_E\\
&+(|12\rangle-|21\rangle)_{AB}|\alpha_7\rangle_E+(|13\rangle-|20\rangle)_{AB}|\alpha_8\rangle_E.\\
\end{array}
\end{equation}
Substituting Eq.(9) into Eq.(6.2), we have
\begin{equation}
\begin{array}{ll}
|\alpha_1\rangle_E=|\alpha_6\rangle_E,
&|\alpha_2\rangle_E=|\alpha_5\rangle_E,\\
|\alpha_3\rangle_E=|\alpha_8\rangle_E, &
|\alpha_4\rangle_E=|\alpha_7\rangle_E.\end{array}
\end{equation}
Hence,  the quantum state $|\Psi\rangle_{ABE}$ can be written as
\begin{equation}
\begin{array}{ll}
&|\Psi\rangle_{ABE}\\
=&(|00\rangle-|33\rangle+|11\rangle-|22\rangle)_{AB}|\alpha_1\rangle_E\\
&+(|01\rangle-|32\rangle+|10\rangle-|23\rangle)_{AB}|\alpha_2\rangle_E\\
&+(|02\rangle-|31\rangle+|13\rangle-|20\rangle)_{AB}|\alpha_3\rangle_E\\
&+(|03\rangle-|30\rangle+|12\rangle-|21\rangle)_{AB}|\alpha_4\rangle_E.\\
\end{array}
\end{equation}
By using Eq.(6.3) one can obtain
\begin{equation}
\begin{array}{ll}|\alpha_1\rangle_E=0, ~~~~~ & ~~~~ |\alpha_3\rangle_E=0.\end{array}
\end{equation}
This leads to
\begin{equation}
\begin{array}{ll}
&|\Psi\rangle_{ABE}\\
=&(|01\rangle-|32\rangle+|10\rangle-|23\rangle)_{AB}|\alpha_2\rangle_E\\
&+(|03\rangle-|30\rangle+|12\rangle-|21\rangle)_{AB}|\alpha_4\rangle_E.\\
\end{array}
\end{equation}

Since the quantum state $|\Psi\rangle_{ABE}$ should satisfy
Eq.(6.4), so there must be
 \begin{equation}|\alpha_4\rangle_E=0.\end{equation}

Thus we arrive at  the conclusion which says
\begin{equation}
\begin{array}{ll}
|\Psi\rangle_{ABE}
=&(|01\rangle-|32\rangle+|10\rangle-|23\rangle)_{AB}|\alpha_2\rangle_E.\\
\end{array}\end{equation}
As a matter of fact, the quantum $|\Psi\rangle_{ABE}$ is just the
quantum state $|\chi\rangle_{AB}|\alpha\rangle_E$. This means  the
proof has been completed.

The above result indicates that when the quantum state
 satisfies Eq.(6), it must be $|\chi\rangle_{AB}|\alpha\rangle_E$. It means that the
quantum channel shared by Alice and Bob  is entirely uncorrelated
with eavesdropper. So it is impossible for the eavesdropper to
obtain  the secret key.

 Therefore,  the legitimate  correspondents, Alice and
Bob can check whether the quantum channel is $ |\chi\rangle_{AB}$ or
not. Each of them choose randomly the operators $\sigma_x, \sigma_z,
\upsilon_x, \upsilon_z$ to measure the entangled states they shared.
After a series of the entangled states have been measured, Alice and
Bob announce the  operators and the measurement outcomes. If Eq.(6)
is satisfied by all measurement outcomes, then the quantum channel
is the entangled quantum state $|\chi\rangle_{AB}$. It means that
the quantum channel shared by Alice and Bob is secure. Otherwise,
they restart the protocol.

  If the quantum channel
 is secure, then two correspondents, Alice
and Bob could use it to distribute quantum keys by the following
steps:

1. Alice and Bob measure their respective particles $A$, $B$ in the
same basis defined in Eq.(2).

2. The quantum channel  $|\chi\rangle_{AB}$  shared by  Alice and
Bob could read
 \begin{equation}
\begin{array}{ll}
~&|\chi\rangle_{AB}\\
=&\frac
{1}{2}(|\psi^-\rangle|\phi^+\rangle+|\phi^-\rangle|\psi^+\rangle
+|\psi^+\rangle|\phi^-\rangle+|\phi^+\rangle|\psi^-\rangle)_{AB}.
\end{array}\end{equation}

In Eq.(16), the four states  $\{|\phi^+\rangle, |\phi^-\rangle,
|\psi^+\rangle, |\psi^-\rangle\}$ could be coded into two bits of
classical information as they are orthogonal to each other. One bit
is used to discriminate the states $|\phi\rangle$ or $|\psi\rangle$
which we call parity bit and the other bit is used to discriminate
the superscripts of the states $|\phi^\pm\rangle$  or
$|\psi^\pm\rangle$  which we call phase bit.

3. From a series of measurement results they choose randomly part of
them
 to reexamine the security of the quantum channel.  If their
results are well correlated, then the quantum channel should be
considered as secure. Otherwise, they should restart the protocol.

4. If the quantum channel is secure by above examination, we could
 use  the parity bits and the phase bit of the measurement
results  as the secret key.

So far, a secret key has been set up between the two correspondents
by using the entangled four-level particles. The security of the
protocol is based on the law  of physics.

 Now we describe  a controlled quantum key distribution protocol by
utilizing three  four-level particles.

Assume that the controller Alice, and two correspondents, Bob and
Charlie, share a quantum channel consisted of a number of the
entangled states of three four-level particles
\begin{equation}
\begin{array}{ll}
&|\chi\rangle_{ABC}\\
=&\frac {1}{2\sqrt 2}(|000\rangle+|011\rangle
+|101\rangle+|110\rangle\\
&~~~~~+|223\rangle+|232\rangle
+|322\rangle+|333\rangle)_{ABC}\\
=&\frac{1}{4}[|\phi^+\rangle_A(|\phi^+\rangle|\phi^+\rangle+|\phi^-\rangle|\phi^-\rangle
\\&~~~~~~~~~~~~+|\psi^+\rangle|\psi^+\rangle+|\psi^-\rangle|\psi^-\rangle)_{BC}\\
&~~+|\phi^-\rangle_A(|\phi^-\rangle|\phi^+\rangle+|\phi^+\rangle|\phi^-\rangle
\\&~~~~~~~~~~~~+|\psi^+\rangle|\psi^-\rangle+|\psi^-\rangle|\psi^+\rangle)_{BC}\\
&~~+|\psi^+\rangle_A(|\psi^+\rangle|\phi^+\rangle+|\phi^+\rangle|\psi^+\rangle
\\&~~~~~~~~~~~~+|\psi^-\rangle|\phi^-\rangle+|\phi^-\rangle|\psi^-\rangle)_{BC}\\
&~~+|\psi^-\rangle_A(|\psi^-\rangle|\phi^+\rangle+|\phi^-\rangle|\psi^+\rangle
\\&~~~~~~~~~~~~+|\psi^+\rangle|\phi^-\rangle+|\phi^+\rangle|\psi^-\rangle)_{BC}].\end{array}\end{equation}
Here the four-level particles $A$, $B$ and $C$  belong to Alice, Bob
and Charlie, respectively.

Now we introduce another two measurement operators
\begin{equation}
\begin{array}{l}
\epsilon_x=|2\rangle\langle 3|+|3\rangle\langle 2|+|0\rangle\langle 1|+|1\rangle\langle 0|,\\
o_z=|3\rangle\langle 3|-|1\rangle\langle 1|+|0\rangle\langle 0|+|2\rangle\langle 2|.\\
\end{array}
\end{equation}
It is easy to prove that $\epsilon_x,
 o_z$  are Hermitian operators and
\begin{equation}\label{4}
\epsilon_x^2=o_z^2=I.
\end{equation}
Therefore, the eigenvalues of the operators $\epsilon_x, o_z$ are 1
or -1.

One can check that $|\chi\rangle_{ABC}$ satisfies
\begin{equation}
\begin{array}{rll}
\sigma_x^A\otimes \sigma_x^B\otimes \sigma_x^C|\chi\rangle_{ABC}&=&|\chi\rangle_{ABC},\\
o_z^A\otimes o_z^B \otimes o_z^C|\chi\rangle_{ABC}&=&|\chi\rangle_{ABC},\\
\epsilon_x^A\otimes\epsilon_x^B\otimes
I^C|\chi\rangle_{ABC}&=&|\chi\rangle_{ABC},\\
I^A\otimes
\epsilon_x^B\otimes\epsilon_x^C|\chi\rangle_{ABC}&=&|\chi\rangle_{ABC}.
\end{array}\end{equation}

Next we prove that if $|\psi\rangle_{ABCE}$  satisfies the following
equation  \addtocounter{equation}{1}
\begin{align}
\sigma_x^A\otimes \sigma_x^B\otimes \sigma_x^C|\psi\rangle_{ABCE}&=|\psi\rangle_{ABCE},\tag{\theequation .1}\\
o_z^A\otimes o_z^B \otimes o_z^C|\psi\rangle_{ABCE}&=|\psi\rangle_{ABCE},\tag{\theequation .2}\\
\epsilon_x^A\otimes\epsilon_x^B\otimes
I^C|\psi\rangle_{ABCE}&=|\psi\rangle_{ABCE},\tag{\theequation .3}\\
I^A\otimes
\epsilon_x^B\otimes\epsilon_x^C|\psi\rangle_{ABCE}&=|\psi\rangle_{ABCE},\tag{\theequation
.4}
\end{align} then
$|\psi\rangle_{ABCE}=|\chi\rangle_{ABC}|\beta\rangle_E$ holds. Here
$E$ still stands for the environment.

The general  formation of the quantum state of the three four-level
particles and the environment should be
\begin{equation}
\begin{array}{ll}
&|\psi\rangle_{ABCE}\\
=&|000\rangle|\beta_{000}\rangle+|001\rangle|\beta_{001}\rangle+|002\rangle|\beta_{002}\rangle+|003\rangle|\beta_{003}\rangle\\
&+|010\rangle|\beta_{010}\rangle+|011\rangle|\beta_{011}\rangle+|012\rangle|\beta_{012}\rangle+|013\rangle|\beta_{013}\rangle\\
&+|020\rangle|\beta_{020}\rangle+|021\rangle|\beta_{021}\rangle+|022\rangle|\beta_{022}\rangle+|023\rangle|\beta_{023}\rangle\\
&+|030\rangle|\beta_{030}\rangle+|031\rangle|\beta_{031}\rangle+|032\rangle|\beta_{032}\rangle+|033\rangle|\beta_{033}\rangle\\
&+|100\rangle|\beta_{100}\rangle+|101\rangle|\beta_{101}\rangle+|102\rangle|\beta_{102}\rangle+|103\rangle|\beta_{103}\rangle\\
&+|110\rangle|\beta_{110}\rangle+|111\rangle|\beta_{111}\rangle+|112\rangle|\beta_{112}\rangle+|113\rangle|\beta_{113}\rangle\\
&+|120\rangle|\beta_{120}\rangle+|121\rangle|\beta_{121}\rangle+|122\rangle|\beta_{122}\rangle+|123\rangle|\beta_{123}\rangle\\
&+|130\rangle|\beta_{130}\rangle+|131\rangle|\beta_{131}\rangle+|132\rangle|\beta_{132}\rangle+|133\rangle|\beta_{133}\rangle\\
&+|200\rangle|\beta_{200}\rangle+|201\rangle|\beta_{201}\rangle+|202\rangle|\beta_{202}\rangle+|203\rangle|\beta_{203}\rangle\\
&+|210\rangle|\beta_{210}\rangle+|211\rangle|\beta_{211}\rangle+|212\rangle|\beta_{212}\rangle+|213\rangle|\beta_{213}\rangle\\
&+|220\rangle|\beta_{220}\rangle+|221\rangle|\beta_{221}\rangle+|222\rangle|\beta_{222}\rangle+|223\rangle|\beta_{223}\rangle\\
&+|230\rangle|\beta_{230}\rangle+|231\rangle|\beta_{231}\rangle+|232\rangle|\beta_{232}\rangle+|233\rangle|\beta_{233}\rangle\\
&+|300\rangle|\beta_{300}\rangle+|301\rangle|\beta_{301}\rangle+|302\rangle|\beta_{302}\rangle+|303\rangle|\beta_{303}\rangle\\
&+|310\rangle|\beta_{310}\rangle+|311\rangle|\beta_{311}\rangle+|312\rangle|\beta_{312}\rangle+|313\rangle|\beta_{313}\rangle\\
&+|320\rangle|\beta_{320}\rangle+|321\rangle|\beta_{321}\rangle+|322\rangle|\beta_{322}\rangle+|323\rangle|\beta_{323}\rangle\\
&+|330\rangle|\beta_{330}\rangle+|331\rangle|\beta_{331}\rangle+|332\rangle|\beta_{332}\rangle+|333\rangle|\beta_{333}\rangle.\\
\end{array}
\end{equation}
Here $|\beta_i\rangle$, $i=000, 001, \cdots, 333$,  are the states
of the environment; $|klm\rangle,$ $k,l,m=0,1,2,3$ denote the states
of the three four-level particles  $A, B, C$.

According to Eq.(21.1), we have
\begin{equation}
\begin{array}{lll}
|\beta_{000}\rangle=|\beta_{333}\rangle,&
|\beta_{001}\rangle=|\beta_{332}\rangle,&
|\beta_{002}\rangle=|\beta_{331}\rangle,\\
|\beta_{003}\rangle=|\beta_{330}\rangle,&
|\beta_{010}\rangle=|\beta_{323}\rangle,&
|\beta_{011}\rangle=|\beta_{322}\rangle,\\
|\beta_{012}\rangle=|\beta_{321}\rangle,&
|\beta_{013}\rangle=|\beta_{320}\rangle,&
|\beta_{020}\rangle=|\beta_{313}\rangle,\\
|\beta_{021}\rangle=|\beta_{312}\rangle,&
|\beta_{022}\rangle=|\beta_{311}\rangle,&
|\beta_{023}\rangle=|\beta_{310}\rangle,\\
|\beta_{030}\rangle=|\beta_{303}\rangle,&
|\beta_{031}\rangle=|\beta_{302}\rangle,&
|\beta_{032}\rangle=|\beta_{301}\rangle,\\
|\beta_{033}\rangle=|\beta_{300}\rangle,&
|\beta_{100}\rangle=|\beta_{233}\rangle,&
|\beta_{101}\rangle=|\beta_{232}\rangle,\\
|\beta_{102}\rangle=|\beta_{231}\rangle,&
|\beta_{103}\rangle=|\beta_{230}\rangle,&
|\beta_{110}\rangle=|\beta_{223}\rangle,\\
|\beta_{111}\rangle=|\beta_{222}\rangle,&
|\beta_{112}\rangle=|\beta_{221}\rangle,&
|\beta_{113}\rangle=|\beta_{220}\rangle,\\
|\beta_{120}\rangle=|\beta_{213}\rangle,&
|\beta_{121}\rangle=|\beta_{212}\rangle,&
|\beta_{122}\rangle=|\beta_{211}\rangle,\\
|\beta_{123}\rangle=|\beta_{210}\rangle,&
|\beta_{130}\rangle=|\beta_{203}\rangle,&
|\beta_{131}\rangle=|\beta_{202}\rangle,\\
|\beta_{132}\rangle=|\beta_{201}\rangle,&
|\beta_{133}\rangle=|\beta_{200}\rangle.&
\end{array}
\end{equation}
So the $|\psi\rangle_{ABCE}$ becomes
\begin{equation}
\begin{array}{ll}
&(|000\rangle+|333\rangle)|\beta_{000}\rangle+(|001\rangle+|332\rangle)|\beta_{001}\rangle\\
&+(|002\rangle+|331\rangle)|\beta_{002}\rangle+(|003\rangle+|330\rangle)|\beta_{003}\rangle\\
&+(|010\rangle+|323\rangle)|\beta_{010}\rangle+(|011\rangle+|322\rangle)|\beta_{011}\rangle\\
&+(|012\rangle+|321\rangle)|\beta_{012}\rangle+(|013\rangle+|320\rangle)|\beta_{013}\rangle\\
&+(|020\rangle+|313\rangle)|\beta_{020}\rangle+(|021\rangle+|312\rangle)|\beta_{021}\rangle\\
&+(|022\rangle+|311\rangle)|\beta_{022}\rangle+(|023\rangle+|310\rangle)|\beta_{023}\rangle\\
&+(|030\rangle+|303\rangle)|\beta_{030}\rangle+(|031\rangle+|302\rangle)|\beta_{031}\rangle\\
&+(|032\rangle+|301\rangle)|\beta_{032}\rangle+(|033\rangle+|300\rangle)|\beta_{033}\rangle\\
&+(|100\rangle+|233\rangle)|\beta_{100}\rangle+(|101\rangle+|232\rangle)|\beta_{101}\rangle\\
&+(|102\rangle+|231\rangle)|\beta_{102}\rangle+(|103\rangle+|230\rangle)|\beta_{103}\rangle\\
&+(|110\rangle+|223\rangle)|\beta_{110}\rangle+(|111\rangle+|222\rangle)|\beta_{111}\rangle\\
&+(|112\rangle+|221\rangle)|\beta_{112}\rangle+(|113\rangle+|220\rangle)|\beta_{113}\rangle\\
&+(|120\rangle+|213\rangle)|\beta_{120}\rangle+(|121\rangle+|212\rangle)|\beta_{121}\rangle\\
&+(|122\rangle+|211\rangle)|\beta_{122}\rangle+(|123\rangle+|210\rangle)|\beta_{123}\rangle\\
&+(|130\rangle+|203\rangle)|\beta_{130}\rangle+(|131\rangle+|202\rangle)|\beta_{131}\rangle\\
&+(|132\rangle+|201\rangle)|\beta_{132}\rangle+(|133\rangle+|200\rangle)|\beta_{133}\rangle.\\
\end{array}\end{equation}

Substituting it into Eq.(21.2), one obtains
\begin{equation}
\begin{array}{l}
|\beta_{001}\rangle=|\beta_{002}\rangle=|\beta_{010}\rangle=|\beta_{012}\rangle=|\beta_{013}\rangle\\
= |\beta_{020}\rangle=|\beta_{021}\rangle= |\beta_{023}\rangle=|\beta_{031}\rangle= |\beta_{032}\rangle\\
=|\beta_{100}\rangle=|\beta_{102}\rangle=|\beta_{103}\rangle= |\beta_{111}\rangle=|\beta_{112}\rangle\\
=|\beta_{120}\rangle=|\beta_{121}\rangle= |\beta_{122}\rangle=|\beta_{123}\rangle= |\beta_{130}\rangle\\
=|\beta_{132}\rangle= |\beta_{133}\rangle=0.
\end{array}\end{equation}
It leads to
\begin{equation}
\begin{array}{ll}
&|\psi\rangle_{ABCE}\\
=&(|000\rangle+|333\rangle)|\beta_{000}\rangle+(|003\rangle+|330\rangle)|\beta_{003}\rangle\\
&+(|011\rangle+|322\rangle)|\beta_{011}\rangle+(|022\rangle+|311\rangle)|\beta_{022}\rangle\\
&+(|030\rangle+|303\rangle)|\beta_{030}\rangle+(|033\rangle+|300\rangle)|\beta_{033}\rangle\\
&+(|101\rangle+|232\rangle)|\beta_{101}\rangle+(|110\rangle+|223\rangle)|\beta_{110}\rangle\\
&+(|113\rangle+|220\rangle)|\beta_{113}\rangle+(|131\rangle+|202\rangle)|\beta_{131}\rangle.\\
\end{array}\end{equation}

By the restriction of Eq.(21.3), we obtain
\begin{equation}
\begin{array}{l}
|\beta_{000}\rangle=|\beta_{110}\rangle,
|\beta_{003}\rangle=|\beta_{113}\rangle,
|\beta_{011}\rangle=|\beta_{101}\rangle,\\
|\beta_{022}\rangle=|\beta_{030}\rangle=|\beta_{033}\rangle=|\beta_{131}\rangle=0.\\
\end{array}\end{equation}
It means
\begin{equation}
\begin{array}{ll}
|\psi\rangle_{ABCE}=&(|110\rangle+|223\rangle+|000\rangle+|333\rangle)|\beta_{000}\rangle\\
&+(|113\rangle+|220\rangle+|003\rangle+|330\rangle)|\beta_{003}\rangle\\
&+(|101\rangle+|232\rangle+|011\rangle+|322\rangle)|\beta_{011}\rangle.
\end{array}\end{equation}
Eq.(21.4) further restricts $|\psi\rangle_{ABCE}$ to be of the form
\begin{equation}
\begin{array}{rl}
|\psi\rangle_{ABCE}=&(|110\rangle+|223\rangle+|000\rangle+|333\rangle\\
~&+|101\rangle+|232\rangle+|011\rangle+|322\rangle)|\beta_{000}\rangle\\
=&|\chi\rangle_{ABC}|\beta\rangle_E.\end{array}
\end{equation}
Thus the proof has been completed.

Based on  the  conclusion above,  Alice,  Bob and Charlie can check
whether the quantum channel is $ |\chi\rangle_{ABC}$ or not. Each of
them selects  randomly the operators $\sigma_x, \epsilon_x, o_z$ and
$I$ to measure the entangled states they shared. After a lot of the
entangled states have been measured, Alice, Bob and Charlie publish
the operators and the measurement outcomes. If Eq.(21) is satisfied
by all measurement outcomes, then the quantum channel is the
entangled quantum state $|\chi\rangle_{ABC}$. It means that the
quantum channel shared by Alice,  Bob and Charlie is entirely
uncorrelated  with eavesdropper. Otherwise, they restart the
protocol. Here we assume that in the process of testing  the
security of the quantum channel,  Alice, Bob and Charlie announce
the true results.

  If the quantum channel
 is secure, then two correspondents, Bob and Charlie,
 could use it to distribute quantum keys controlled by Alice.
The details are as follows:

1. Alice, Bob and Charlie measure the rest of their respective
particles $A$, $B$ and $C$ in the same basis defined in Eq.(2) and
record measurement results.

2. Evidently, the measurement results are correlated according to
Eq.(17). Hence the measurement results is a secret key shared by
Alice, Bob and Charlie. If the controller Alice permits Bob and
Charlie to create a secret key, she should tell Bob and Charlie her
measurement outcomes. With the message of Alice's measurement
outcomes and Bob's measurement results, Bob can deduce  Charlie's
measurement results.  At the  same time, with the message of Alice's
measurement outcomes and Charlie's measurement results, Charlie can
obtain Bob's measurement results. So with Alice's permission Bob and
Charlie can create a secret key based on the measurement results.
However, if Alice does not announce her measurement outcomes, there
is no way for Bob and Charlie to create a secret key by using the
quantum channel $|\chi\rangle_{ABC}$. This is just a controlled
quantum key distribution.  In other words, the quantum  key
distribution between Bob and Charlie is controlled by Alice. As a
matter of fact, this controlled quantum key distribution protocol is
just  a secret sharing protocol \cite{HBB, YanGao}.

In summary, we  proposed a  quantum key distribution protocol based
on four-level particle entanglement and described a controlled
quantum key distribution protocol by utilizing three four-level
particles. The security of the two protocols are guaranteed by the
law of quantum physics.

\vspace{0.5cm}

{\noindent\bf Acknowledgments}\\[0.2cm]

This work was supported by the National Natural Science Foundation
of China under Grant No: 10971247, Hebei Natural Science Foundation
of China under Grant No: F2009000311.


\end{document}